# Self-Diagnosis, Scaffolding and Transfer: A Tale of Two Problems

Andrew Mason[1], Elisheva Cohen[2], Chandralekha Singh[1], and Edit Yerushalmi[2]

[1]Department of Physics and Astronomy, University of Pittsburgh, Pittsburgh, PA 15213, USA
[2]Department of Science Teaching, Weizmann Institute of Science, Rehovot, Israel

**Abstract.** Helping students learn from their own mistakes can help them develop habits of mind while learning physics content. Based upon cognitive apprenticeship model, we asked students to self-diagnose their mistakes and learn from reflecting on their problem solution. Varying levels of scaffolding support were provided to students in different groups to diagnose their errors on two context-rich problems that students originally solved in recitation quizzes. Here, we discuss students' cognitive engagement in the two self-diagnosis activities and transfer tasks with different scaffolds.

## INTRODUCTION

Previously we described a formative assessment task in which in the session following a quiz the students self-diagnosed their problem solutions with different levels of external support. The external instructions and resources provided in our studies were the following [1-4]. One group (B, N=31) used both an outline of the solution by the TA and a rubric reflecting general problem solving steps (such as "problem description", "plan", "evaluation") common to several problem solving strategies described in the research literature. Another group (C, N=28) received a detailed worked out example; and the last group (D, N=25) received the final answer and students in group D were allowed to use their notes and textbooks. Group A (two recitation sections, N=87) functioned as a control group. This group could discuss the solution for the quiz with the TA. No attempt was made to prompt students to self-diagnose their solutions.

The first study [1-3] involved an unusually difficult problem (termed quiz 6) and examined transfer to a midterm problem identified in an independent study [5] as a far transfer. For quiz 6, the pre (quiz) and post (midterm) problems involved Newton's 2$^{nd}$ Law in a non-equilibrium situation with centripetal acceleration, combined with conservation of energy. The pre problem dealt with the normal force on a rollercoaster passenger at the top of a circular bump; the post problem asked students to solve for the tension in a rope swing at the point in its motion where the tension was greatest when a person is swinging. The level of innovation required as well as the distance of transfer may have contributed to the relatively poor performance of all students on pre and post problems.

In a companion paper [4], we examined an easier problem (termed quiz 7) and a closer transfer. For quiz 7, the pre and post problems involved the conservation of mechanical energy and the conservation of momentum principles. The quiz problem required calculating the height a person will reach jumping on a skateboard and climbing up a hill. The post required calculating the height a person will reach jumping from a dinosaur on a cart and climbing up a hill.

In both studies we looked at students' performance on two aspects: the physics involved in the solution and diagnosis, and the communication of the solution and diagnosis. In this paper, we consider the results of both studies with respect to the physics aspect.

We first present our expectations for inter-group as well as intra-group comparison. Regarding inter-group comparison, we expect that when more external support will be provided, students will perform better self-diagnosis. Regarding intra-group comparison, we expect a successful self-diagnosis task will help those students who have poor reflective study habits to change their reflective behavior in the context of problem solving. Thus we expect the intervention to reduce the gap between the low and high achievers.

Bransford and Schwartz theorized that the preparation for future learning (PFL) and transfer of knowledge from the situation in which it was acquired to new situations is optimal if instruction incorporates both the elements of innovation and efficiency [6]. In their model, efficiency and innovation are two orthogonal coordinates. If instruction only focuses on efficiency, the cognitive engagement and processing by the students will be diminished and they will not develop the ability to transfer the acquired knowledge to new situations. Similarly, if the instruction is solely focused on innovation, students may struggle to connect what they are learning with their prior knowledge so that learning and transfer will be inhibited. They propose that the transfer will be enhanced if the instructional scaffolding focuses on moving along a diagonal trajectory in the two dimensional space of innovation and efficiency [6]. We translate their theory to our context in the following manner: instruction that makes students solve many problems focuses on moving them along the efficiency coordinate. Instruction that establishes in students reflective study habits moves them along the innovation coordinate. Self-diagnosis (SD) task in its best allow students both to extend their repertoire of solution schemes as well as to develop study habits. Thus, we expect a successful SD task to reduce the gap between low and high achievers.

Yet, the SD task might not be a successful one. Accordingly, we define three possible types of a SD intervention:

The first type we term "weak". A weak intervention does not affect the gap between low and high achievers. The high achieving students display their natural diagnostic tendency, while the low achievers do not have sufficient guidance to reflect, diagnose and learn from the activities. In this case we expect positive correlations between students' quiz score (pre), the self-diagnosis (SD) score, and the midterm score (that serves as "post" or transfer task).

The second type of intervention we term "superficial". The intervention will be "superficial" if low-achieving students score high on the diagnosis of their mistakes, but this diagnosis is not really meaningful. These students do not accommodate their knowledge by doing the diagnosis; hence it does not affect their achievement later on. This might happen if students use the scaffolding tools to realize principles/concepts that were not invoked or applied correctly, but do not learn why the way the principles/concepts were applied was wrong and how they should be applied correctly. Since most students in our research did not provide such explanations, the self-diagnosis score in itself does not allow us to know if the diagnosis was accompanied by a superficial or meaningful learning process. If the superficial effect is dominant in an intervention, we expect no significant correlation between the scores on pre problem and its SD since the SD of low achievers might be reasonable. But we expect no correlation between the scores on SD and the post problem, since superficial SD does not allow for transfer to occur. We expect positive correlations between the pre and the post problem in this case, since the situation is the same as if the students haven't diagnosed themselves.

**TABLE 1.** Expected correlations between different variables. "N/S" stands for not significant.

| Type of correlation | Control Group | Type of Intervention | | |
|---|---|---|---|---|
| | | Weak | Superficial | Meaningful |
| pre vs. SD | N/A | + | N/S | N/S |
| SD vs. post | N/A | + | N/S | + |
| Pre vs. post | + | + | + | N/S |

The third possible type of intervention we termed as "meaningful", namely one that brings even the less achieving students to perform a meaningful diagnosis of their mistakes, and affects their achievement later on. According to Chi [7], we expect the intervention to be meaningful if two things happen: the student a) compares two textual artifacts, the sample solution and their solution, and realizes *omissions* (i.e. differences that are significant to finding the right solution), and b) acknowledges *violations*, i.e. conflicts between a text sentence in the sample solution and a belief that is embedded in the mental model of the student, instigating the self-repair of their mental model.

If this effect is dominant in an intervention, we expect non-significant correlation between the pre problem and its SD, since the low achievers will reduce the gap between them and the better students. Also, we expect a positive correlation between the SD and the post problem, since this SD does affect the low achievers' performance on the transfer problems. Also, we expect non-significant correlations between the pre and post problems, since the formerly weak students have actually learned from the self-diagnosis and are likely to perform better later.

For the control groups, we expect to get positive correlations between the pre and the post, since no intervention intended to reduce the gaps between low and high achievers took place. Assuming one of these three effects is dominant in an intervention; these intra-group expectations are summed up in Table 1.

In the above framework, if we consider a case in which the transfer problem is a far transfer problem, we can not predict the pre-post correlations. However, if the pre post correlation for the control group is positive, it would be reasonable to assume the pre-post correlations for the intervention groups would be similar to the expected correlations in table 1.

## Findings – inter-group comparisons

In both quiz problems (pre), students' initial quiz performance was relatively poor, even though they showed improvement on second quiz (mean physics score over all groups ~ 0.38 for quiz 6 vs. ~0.46 for quiz 7). The self-diagnosis (SD) performance in quiz 6 and quiz 7 were dissimilar. Table 2 shows that for quiz 6, there was a definite difference between groups on self-diagnosing physics mistakes based on external support provided (p value < 0.05). This difference was not there for quiz 7. Group D particularly stood out as able to self-diagnose with only textbook and notes.

**TABLE 2**. SD Physics grades - two studies

|        |           | Group B | Group C | Group D |
|--------|-----------|---------|---------|---------|
| Quiz 6 | Mean      | 0.73    | 0.57    | 0.24    |
|        | Std. Err. | 0.049   | 0.051   | 0.055   |
| Quiz 7 | Mean      | 0.56    | 0.62    | 0.61    |
|        | Std. Err. | 0.056   | 0.06    | 0.065   |

Group D students received merely the textbook and notes when self-diagnosing their solutions. These resources proved as adequate for the level of difficulty in quiz 7 but not for quiz 6. How is it possible?

The fact that students who used their notes and textbook did well on self-diagnosis of quiz 7 suggests that for this problem, the textbooks and notes included solution for similar problems. Quiz 7 involved conservation of mechanical energy and conservation of momentum. Indeed, examples of solved problems which involve momentum conservation with completely inelastic collision and the conservation of mechanical energy exist in the textbook students used and the instructor also presented related sample solutions in class. Moreover, one solved example in the textbook was about the ballistic pendulum. Although the surface features of the quiz problem which involves a person jumping on a skateboard and climbing up a hill are different from those of the ballistic pendulum problem, both solutions involve in a similar manner the principles of momentum and mechanical energy conservation. In quiz 6, however, the textbooks and notes did not contain solutions to problems similar to the quiz problem.

These results suggest, first, that when diagnosing their problem solution for quiz 7 students in group D were able to make use of solutions for problems sharing similar solution procedure while differing in surface features; and second, that without an access to such a repertoire of related problem solutions, they were not able to self-diagnose their problem solution. Thus, such repertoire brings the self-diagnosis task within the zone of proximal development for group D.

The comparison of the post (midterm exam) scores for quiz 6 and quiz 7 shows that the midterm physics scores (post) improved significantly in the second study compared to the first study for all groups (0.44 in midterm II and 0.61 in midterm III). Despite the difference in SD performance, the groups did not differ significantly from each other in the post performance for quiz 6. Group D fared better than group C on the midterm (C-0.33, D-0.47, p-value=0.07) despite the fact that group C got the complete solution. The fact that the SD grade did not predict the post performance, suggests the SD for quiz 6 did not reflect a meaningful self-diagnosis. In quiz 7 there was no significant difference between groups both on the SD as well as on the midterm performance. Yet, this does not imply a more meaningful diagnosis in quiz 7. The intra-group correlations can shed light on that.

## Findings – intra-group comparisons

Table 3 shows the observed correlations in the three interventions related to the difficult quiz 6. The results are confusing: on one hand in all intervention groups there are no significant correlations between the pre vs. post whereas there is a positive correlation between these two variables in the control group, suggesting that the intervention indeed has reduced the gaps between low and high achievers. On the other hand there are no significant intra-group correlations for any of the intervention groups regarding the SD vs. post, implying that we cannot simply declare any of the three interventions as completely superficial or meaningful (compare Tables 1).

To explain this, oOne possible explanation is that the non-significant correlation between the SD vs. post problems is due to the fact that the post problem was a far transfer to the quiz. Another explanation is that some students may have performed meaningful self-diagnosis that does not show in their self-diagnosis grades.

A possible support for this interpretation is in the inter-groups comparison. The support for self-diagnosis was not adequate between groups, and indeed group D's average self-diagnosis score in quiz 6 was low (0.24) compared to groups B (0.73) and C (0.57). However, the midterm (post) performance of group D (mean=0.47) was comparable to group B (mean=0.52) (group C performed worse; mean=0.33)) It is possible that struggling with the diagnostic activity without a sample solution may have stimulated out of class diagnosis. Frustrated by two failed attempts at the problem (quiz and diagnosis), group D students may have diagnosed their solutions after the in-class activity was completed, with the sample solution that all students received after the fact.

**TABLE 3.** Correlations physics aspect - two studies

|        | Type of correlation | Control group | Intervention D | C | B |
|--------|---------------------|---------------|---|---|---|
| quiz 6 | Pre vs. SD          | N/A | N/S | N/S | N/S |
|        | SD vs. post         | N/A | N/S | N/S | N/S |
|        | Pre vs. post        | +   | N/S | N/S | N/S |
| quiz 7 | Pre vs. SD          | N/A | N/S | N/S | N/S |
|        | SD vs. post         | N/A | +   | N/S | N/S |
|        | Pre vs. post        | +   | N/S | N/S | N/S |

Table 3 shows also the correlations found in the second study related to the easier quiz 7. In this case, groups B and C have no significant correlations between any two variables similar to quiz 6. However, there is a positive correlation between SD vs. post (but there is no correlation between pre vs. post and pre vs. SD) for group D (correlation=0.53, p value<0.05). Thus, group D seems to correspond to a "meaningful" intervention. Group D did in fact perform at least as well as the other groups on both the self-diagnosis and post problem [4].

## Discussion

Inter-group findings: For the SD of quiz 6 we found that the greater the external support is, the better the SD is. However, in quiz 7 we found no differences between the SD of the different groups. We can explain this by saying that if a problem is a difficult one, whose solution (or similar solution) is hard to find in textbooks, students need much more support in order to self-diagnose their mistakes. However, if the problem is a conventional one, and similar problems appear in textbooks and notes, students may be able to do the hard work by themselves. Moreover, the SD grade does not necessarily imply meaningful SD-for example, the SD average grade for group B in quiz 6 was much better than that for the other groups, but their post average was comparable to groups B and C.

Intra-groups findings: As mentioned above, the SD grade does not necessarily imply meaningful SD-this can also be seen via the intra group correlations-for all groups there is a non significant correlation between pre and SD of quiz 6. This may imply that the lower achieving students reduced the gap between them and the better students. However, we found no correlation between SD and post for all groups. This means that the SD was not meaningful. The same is true for quiz 7, expect for the positive correlation for group D between SD and post. So what allowed group D students to perform a meaningful diagnosis in this case? A possible answer might reside in focusing on the type of mistakes students made. We differentiated between mistakes concerning invoking the principles needed to solve the problem, and mistakes in the careful application of this principle [2,4]. Comparison of the self-diagnosis on quiz 6 and quiz 7 shows that on quiz 6, a majority of students who had difficulty in invoking both physics principles (conservation of mechanical energy and Newton's second law) were able to self-diagnose their mistakes in invoking, but in quiz 7, a large number of students were not able to self-diagnose their mistakes in invoking one of the two principles. The difficulty in self-diagnosing the mistake in invoking the momentum conservation principle was common for quiz 7 among students of all groups. Comparison of students' mistakes in applying physics principles in quiz 6 and quiz 7 shows that approximately 60% of the students (including all groups) were unable to apply the physics principles correctly even if they invoked it in quiz 6 whereas for quiz 7, more than 90% of the students who invoked a physics principle were able to apply it correctly.

The situation in quiz 6 strengthens the assumption that for groups C and B only superficial characteristics of the SD could be analyzed by the researchers, as it was very easy for these students to only compare the sample solution with their solution and realize they differ in the principles that should have been invoked.

We suspect that the difference between group D and groups B and C may be that the latter two groups got much more support to do the SD. In contrast to what one might assume, this support might actually have resulted in a superficial SD performance by a majority of students in these groups. Using the terminology defined by Chi [7]: they merely noticed omissions and external differences between their own solution and the instructor's solution, but the cognitive engagement was very low and there was little opportunity for conflict between their mental model and the instructor's model, thus not leading to self-repair in students' mental model [7].

Finally, we hypothesize that for a conventional problem, students must complete two stages to achieve a meaningful intervention: realizing their omissions and realizing the conflict between their thought and the text. However, the second stage might only be reflected and observed in the SD of students who received the minimal support. This is because students who compare their solutions to the sample solution might only state: "I did not do this equation". From this statement we (the researchers) would not be able to differentiate between students who could self-repair their mental model, and those who could not.

On the contrary, for students in group D, who had to work harder to find a solution to a problem that is similar to the problem they are trying to diagnose, the grade indicated a meaningful diagnosis.

## ACKNOWLEDGMENTS

We thank ISF 1283/05 and NSF 0442087 for support.